\documentclass[prd, aps, nofootinbib, preprintnumbers, showpacs, superscriptaddress, twocolumn]{revtex4}

\usepackage{graphicx,epsfig}% Include figure files
\usepackage{bm}
\usepackage{amsmath}
\usepackage{amsfonts}

\begin{document}

\title{Final spin of a coalescing black-hole binary: an Effective-One-Body approach}

\author{Thibault Damour}
\affiliation{Institut des Hautes Etudes Scientifiques, 35 route de Chartres, 91440 Bures-sur-Yvette, France}

\author{Alessandro Nagar}
\affiliation{Dipartimento di Fisica, Politecnico di Torino, Corso Duca degli Abruzzi 24,
              10129 Torino, Italy and INFN, sez. di Torino, Via P.~Giuria 1,
              Torino, Italy}

\begin{abstract}
We update the analytical estimate of the final spin of a coalescing black-hole binary
derived within the Effective-One-Body (EOB) approach. We consider unequal-mass non-spinning
black-hole binaries. It is found that a more complete account of relevant physical effects
(higher post-Newtonian accuracy, ringdown losses) allows the {\it analytical} EOB estimate to
``converge towards'' the recently obtained {\it numerical} results within $2\%$.
This agreement illustrates the ability of the EOB approach to capture the essential physics 
of coalescing black-hole binaries. Our analytical approach allows one
to estimate the final spin of the black hole formed by coalescing binaries in a mass
range ( $\nu=m_1m_2/(m_1+m_2)^2 < 0.16 $) which is not presently covered by numerical
simulations.
\end{abstract}

\date{\today}

\pacs{
%04.00.00 General relativity and gravitation  (see also 95.30.Sf in astronomy) 
% ... ... Special relativity, see 03.30.+p 
%04.20.-q Classical general relativity  (see also 02.40.-k Geometry,  differential geometry, and  topology)
%04.20.Cv Fundamental problems and general formalism 
%04.20.Dw Singularities and cosmic censorship 
%04.20.Ex Initial value problem, existence and uniqueness of solutions 
%04.20.Fy Canonical formalism, Lagrangians, and variational principles 
%04.20.Gz Spacetime topology, causal structure, spinor structure 
%04.20.Ha Asymptotic structure 
%04.20.Jb Exact solutions 
%04.25.-g Approximation methods; equations of motion 
%04.25.Dm Numerical relativity 
04.25.Nx, %Post-Newtonian approximation; perturbation theory; related approximations 
04.30.-w, %Gravitational waves: theory 
04.30.Db %Wave generation and sources 
%04.30.Nk Wave propagation and interactions 
%04.40.-b Self-gravitating systems; continuous media and classical fields in curved spacetime 
%04.40.Dg Relativistic stars: structure, stability, and oscillations  (see also 97.60.-s Late  stages of stellar evolution)
%04.40.Nr Einstein-Maxwell spacetimes, spacetimes with fluids, radiation or classical fields 
%04.50.+h Gravity in more than four dimensions, Kaluza-Klein theory, unified field theories; alternative theories of gravity  (see also 11.25.Mj Compactification and four-dimensional models)
%04.60.-m Quantum gravity 
%04.60.Ds Canonical quantization 
%04.60.Gw Covariant and sum-over-histories quantization 
%04.60.Kz Lower dimensional models; minisuperspace models 
%04.60.Nc Lattice and discrete methods 
%04.60.Pp Loop quantum gravity, quantum geometry, spin foams 
%04.62.+v Quantum field theory in curved spacetime 
%04.65.+e Supergravity  (see also 12.60.Jv Supersymmetric models)
%04.70.-s Physics of black holes  (see also 97.60.Lf-in astronomy)
%04.70.Bw Classical black holes 
%04.70.Dy Quantum aspects of black holes, evaporation, thermodynamics 
%04.80.-y Experimental studies of gravity 
%04.80.Cc Experimental tests of gravitational theories 
%04.80.Nn Gravitational wave detectors and experiments  (see also 95.55.Ym-in astronomy)
%04.90.+e Other topics in general relativity and gravitation (restricted to new topics in section 04) 
}
\maketitle

%------------------------------
\section{Introduction}
\label{sec:intro}
%------------------------------1
The dimensionless spin parameter $\hat{a}=a/M=J/M^2$
of the black hole formed by the coalescence of a black-hole binary is a
complicated function of the initial masses and spins of the two constituent black holes.
[We consider inspiralling systems circularized by radiation reaction].
The function  $\hat{a}(m_1,m_2, {\bf S}_1, {\bf S}_2)$ is a useful
diagnostic for comparing analytical approaches to the dynamics of
coalescing binaries to the results of three-dimensional numerical 
simulations. Here we shall consider the simple case of {\it non-spinning} binaries,
where  $\hat{a}$ becomes simply a function of the symmetric mass ratio
 $\nu=m_1m_2/(m_1+m_2)^2$ (which satisfies $ 0 \leq \nu \leq 1/4$).
 
 As far as we know, the first estimate of $\hat{a}$ was made in 
2000 \cite{Buonanno:2000ef},
on the basis of a new analytical approach to the general relativistic two-body 
dynamics, the {\it Effective-One-Body} (EOB) approach \cite{Buonanno:1998gg},
at a time where  no reliable numerical simulations of coalescing 
black hole binaries were yet available. This estimate was $\hat{a}= 0.795$
for $\nu = 1/4$ (i.e. for the equal mass case), and was based on estimating
the ratio $J/M^2$ at a ``matching radius''  $r_{\rm match} \simeq 2.85M$ where the two-black-hole
system was replaced by a unique, ringing black hole. This estimate used a 2.5 post-Newtonian
(PN) accurate description of the dynamics down to $r_{\rm match}$, and neglected
the energy and angular momentum losses during the ringdown. In 2001, a combination
of full numerical simulations with a ``close-limit'' approximation \cite{Price:1994pm}
describing the ringing final black hole  led to a similar estimate,
namely $\hat{a} \simeq 0.8$ for $\nu = 1/4$~\cite{Baker:2001nu}. This estimate was
revised downwards in 2002, to $\hat{a} \simeq 0.7$ \cite{Baker:2002qf}, when a better
({\it Lazarus}-based) way of bridging the far- and close-limit approaches
indicated that the angular momentum losses were larger than previously
estimated. Indeed, Ref.~\cite{Baker:2001nu} had estimated the total angular momentum loss
after crossing the Last Stable Orbit (LSO)~\footnote{Also known as the
innermost stable circular orbit (ISCO).} to be around $2\%$,
and the corresponding energy loss to be around $3\%$. [The EOB approach had
estimated an energy loss beyond the LSO of about $1.4 \%$, half of it emitted
during the plunge and the other half during the ringdown~\cite{Buonanno:2000jz}].
By contrast, Ref.~\cite{Baker:2002qf} estimated the angular momentum loss
to be around $12\%$. More recently, an update of the analytical EOB
estimate~\cite{Buonanno:2005xu} using the now available 3.5PN accurate
description of the two-body dynamics showed that  increasing
the PN accuracy of the dynamics (from 2.5PN to 3.5PN) had the effect
of decreasing the final spin parameter from $\hat{a}= 0.795$ to $\hat{a}= 0.77$
(again for $\nu = 1/4$, and again when  neglecting the angular
momentum loss below the matching radius $r_{\rm match}$).

The estimates just recalled belong to a ``prehistoric'' era where, for a variety of reasons,
numerical simulations did not exhibit a very convincing convergence among
themselves, nor towards analytical results (see, however,~\cite{Baker:2002qf}
and~\cite{Damour:2002qh}). This era has recently ended thanks to remarkable
breakthroughs in numerical relativity. Different groups have finally succeded
in numerically simulating the merger of two black holes 
of comparable masses $m_1$ and $m_2$, possibly with spin, and their results
exhibit convincing internal convergence, and a nice mutual consistency~\cite{Pretorius:2005gq,Campanelli:2005dd,Diener:2005mg,Campanelli:2006gf,
Baker:2006yw,Baker:2007fb,Campanelli:2006uy,Gonzalez:2006md,
Koppitz:2007ev,Thornburg:2007hu,Husa:2007rh,Husa:2007hp}.
In the particular case of non-spinning black holes with equal masses,
$m_1=m_2$, different groups now agree on the value $\hat{a}\simeq 0.69$ 
(which is within the span of the ``prehistoric'' estimates recalled above).
The most extensive analysis of the final angular momentum of coalescing
black holes (in the non-spinning case) to date has been carried out by 
Gonzalez et al.~\cite{Gonzalez:2006md}: they have considered a large
sample of unequal-mass systems, corresponding to a symmetric
mass ratio $\nu$ varying in the range $0.1613 \leq \nu \leq 1/4$, 
and have accurately determined the variation of $\hat{a}$ with $\nu$ within this
range. The aim of the present paper is to generalize and update the 
analytical EOB estimates \cite{Buonanno:2000ef,Buonanno:2005xu} of $\hat{a}(\nu)$
recalled above,  both by explicitly considering general values of $\nu$ and by 
improving the previous EOB treatments  of the physical 
ingredients which are crucial in determining the value of $\hat{a}$. Our motivation
for this study is two-fold: on the one hand, we wish to see to what extent numerical
results can be reproduced by analytical (EOB) estimates, and on the other hand
we wish to understand, on this example, what are the physical 
ingredients which are crucial in accurately determining the plunge,
merger and ringdown dynamics of coalescing black holes.

Let us recall that the EOB approach to the general relativistic two-body 
dynamics is a {\it non-perturbatively resummed}  analytic technique 
which has been developed in 
Refs.~\cite{Buonanno:1998gg,Buonanno:2000ef,Damour:2000we,Damour:2001tu,Buonanno:2005xu,Damour:2006tr}.
This technique uses, as basic input,  the results of PN theory,
such as: (i) PN-expanded equations of motion for two point-like bodies,
 (ii) PN-expanded radiative multipole moments, and (iii) PN-expanded
energy and angular momentum fluxes at infinity. For the moment, the
most accurate such results are the 3PN conservative dynamics 
\cite{Damour:2001bu,Blanchet:2003gy}, and the 3.5PN energy 
flux~\cite{Blanchet:2001aw,Blanchet:2004bb,Blanchet:2004ek}. Then
the EOB approach ``packages''
this PN-expanded information in special {\it resummed} forms
which extend the validity of the PN results beyond the 
expected weak-field-slow-velocity regime into (part of) the 
strong-field-fast-motion regime. The aim being to use the EOB approach
for analytically describing the last inspiralling orbits, the transition from inspiral 
to plunge, and the plunge itself down to a ``matching radius''  $r_{\rm match}  $
small enough to allow one to match there the plunge waveform to a ringdown one.

The basic new ingredient used below to improve the previous EOB estimates of  $\hat{a}$
is an approximate treatment of the energy
and angular momentum losses during ringdown. These losses were neglected
in~\cite{Buonanno:2000ef,Buonanno:2005xu}. Our approximation will consist in
estimating these losses by {\it rescaling} (proportionally to $\nu^2$)
the losses obtained by numerically studying the test-mass limit 
\`a la Regge-Wheeler-Zerilli~\cite{Nagar:2006xv,Damour07a}.
We shall also study the effect of the radiation reaction force during the plunge.
Instead of using (as in \cite{Buonanno:2000ef}) a naive analytic continuation
of the radiation reaction appropriate to the inspiral phase, we shall also use
another Pad\'e resummed radiation reaction force, which does not assume
the validity of Kepler's law $\Omega^{2} r^3 = $const. during the 
plunge~\cite{Damour:2006tr}. Indeed, the recent Regge-Wheeler-Zerilli-like
study of the waveform emitted by plunging test-masses~\cite{Nagar:2006xv,Damour07a}
has shown that this modified radiation reaction stays closer to the ``exact''
gravitational wave angular momentum flux computed  \`a la Regge-Wheeler-Zerilli.

The paper is organized as follows. In Sec.~\ref{sec:equations} we review the
non-perturbative construction of the two-body dynamics incorporating
radiation reaction effects while we devote Sec.~\ref{sec:results} to the 
presentation of our results. Some conclusions are presented in Sec.~\ref{sec:conclusions}.
We use geometric units $G=c=1$.
%-----------------------------
\section{Equations}
\label{sec:equations}
%-----------------------------
In this section we recall the non-perturbative construction of the
two-body dynamics including a radiation reaction force.
We take advantage of the most complete PN results, i.e. we work
at 3PN for the conservative part of the dynamics and at 3.5PN
for the radiation damping.
In the EOB framework, the complicated PN-expanded 
relative dynamics (in the center
of mass frame) of the binary system of masses $m_1$ and $m_2$  
is mapped\footnote{modulo some quartic-in-momenta additional terms; see below.}
 into the simpler geodesic dynamics of  a particle of mass 
$\mu=m_1m_2/(m_1+m_2)$ moving in some effective background geometry 
(in  Schwarzschild gauge) 
\begin{equation}
\label{eq:eff_metric}
ds^2 = -A(r)dt^2 + B(r)dr^2 + r^2\left(d\theta^2 + \sin^2\theta d\varphi^2\right) \ .
\end{equation}
Here, and below, we work with the dimensionless reduced variables $r=R/M$ and $t=T/M$, with $M=m_1+m_2$; 
($r,\theta,\varphi$)  are polar coordinates in the {\it effective} problem 
that describe the relative motion.The coefficients of the effective metric 
at the 3PN approximation~\cite{Damour:2000we} read
\begin{align}
\label{eq:A}
A^{\rm 3PN}(r) &\equiv 1-\dfrac{2}{r} + \dfrac{2\nu}{r^3} +\left(\dfrac{94}{3}-\dfrac{41}{32}\pi^2\right)\dfrac{\nu}{r^4} \ , \\
\label{eq:D}
(BA)^{\rm 3PN}(r)& \equiv D^{\rm 3PN}(r)\equiv 1-\dfrac{6\nu}{r^2} + 2(3\nu-26)\dfrac{\nu}{r^3} \ .
\end{align}
Note that we work with the PN expansion of the quantity $D(r) \equiv B(r) A(r)$,
rather than with the $g_{rr}$ metric coefficient 
$B(r) = D(r)/A(r)$. [Recall that $B(r)$ is equal to $1/A(r)$ in the 
Schwarzschild case (which corresponds to the test-mass limit $\nu \to 0$).]
 
\begin{figure}[t]
\begin{center}
\includegraphics[width=85 mm, height=75 mm]{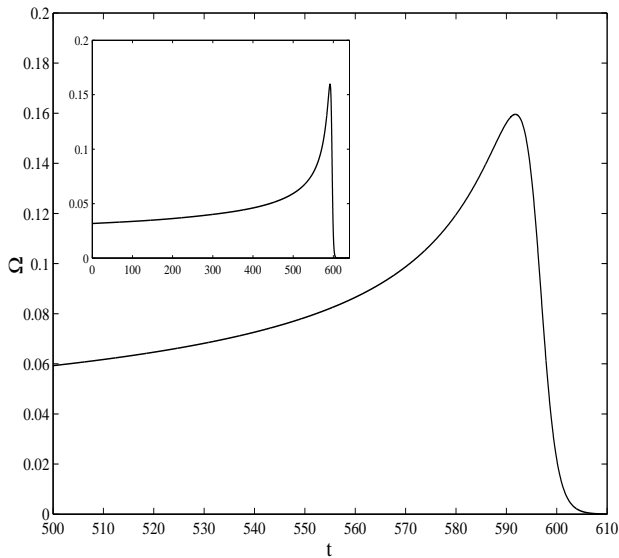}
\caption{\label{fig:fig1}Time evolution of the orbital frequency 
$\Omega$ for $\nu=1/4$: the maximum occurs near the $\nu$-deformed
light ring at $r_{\rm LR}(1/4)\approx 2.316$.}
\end{center}
\end{figure}

Though the EOB packaging of the complicated original PN dynamics into the
much simpler metric coefficients $A(r), D(r)$ already represents an
efficient resummation of PN-expanded results, it is quite useful, 
especially at 3PN, to further resum the `packages' $A(r), D(r)$. 
Indeed, a general feature of the EOB philosophy is to smoothly connect 
the EOB structures to their $\nu \to 0$ 
limit~\cite{Damour:2000we,Damour:2001tu}. Here, 
this can be done by replacing 
the (Taylor expanded) metric coefficients given by 
Eqs.~(\ref{eq:A})-(\ref{eq:D}) with suitable Pad\'e approximants.
The simplest, and most robust, choices consist in using, as metric functions,
the following definitions
$A(r)\equiv P^1_3[A^{\rm 3PN}]$ 
and $D(r)\equiv P^0_{3}[D^{\rm 3PN}]$
\footnote{We recall that  $P^0_3[D^{\rm 3PN}]=1/(1+6\nu u^2
  -2(3\nu-26)\nu u^3)$ where $u=1/r$, while $P^1_{3}[A^{\rm 3PN}]$ is explicitly 
given in Eq.~(4.40c) of Ref.~\cite{Damour:2000we}.  }. 
These Pad\'e approximants are used to ensure the following two facts: 
(i) the function $A$ has a simple zero for a positive value of $r$ 
(like $A_{\rm Schw}=1-2/r$) and (ii)  the function $D$ stays positive while 
$r$ decreases (like $D_{\rm Schw}=1$).
The Pad\'e resummation of $A$ is useful for ensuring the existence and 
$\nu$-continuity of a last stable orbit (LSO), as well as the existence
and $\nu$-continuity of a last unstable orbit, i.e. of a $\nu$-deformed
analog of a light ring (LR). We recall that the LR corresponds to the
circular orbit of a massless particle, or of an extremely relativistic
massive particle, and is technically defined by looking for the maximum
of $A(r)/r^2$, i.e. by solving $ (d/dr) (A(r)/r^2) =0$.
The Pad\'e resummation of $D$ is useful  to ensure that
the orbital frequency $\Omega=d\varphi/dt$ has a clear maximum at
(approximately) the EOB $\nu$-deformed light ring and then drops to zero
(cf Fig.~\ref{fig:fig1} below for $\nu=1/4$).

In the EOB approach one splits the general relativistic relative dynamics of 
a binary system  into a conservative part, determined by the EOB
Hamiltonian defined below, and a non-conservative part related to the loss of angular 
momentum through gravitational radiation. The EOB  Hamiltonian (divided
by $\mu$) is given by
\begin{equation}
\hat{H} \equiv \dfrac{1}{\nu}\sqrt{1+2\nu\left(\hat{H}_{\rm eff}-1\right)}
\end{equation}
where $\hat{H}_{\rm eff}$ denotes the so-called 
``effective  Hamiltonian'' (describing the geodesic dynamics
of the ``effective'' test-mass $\mu$), originally written as~\cite{Damour:2000we}
\begin{equation}
\hat{H}_{\rm eff} = \sqrt{
    A\left(1+\dfrac{p_\varphi^2}{r^2}+\dfrac{p_r^2}{B} +z_3 \dfrac{p_r^4}{r^2}
    \right)} \ .
\end{equation}
Here $z_3=2\nu(4-3\nu)$,
 $\hat{H}\equiv H/\mu$, $\hat{H}_{\rm eff}\equiv H_{\rm eff}/\mu$, 
$p_{\varphi}\equiv P_{\varphi}/(\mu M)$, $r=R/M$ and $p_r$
denotes the conjugate momentum to $r$.

Following what we did in the test-mass limit case~\cite{Nagar:2006xv,Damour07a}, 
the relative dynamics is somewhat more conveniently described by replacing the 
Schwarzschild-like radial variable $r$ by the EOB generalization of the 
Regge-Wheeler tortoise coordinate $r_*$, defined by integrating
\begin{equation}
\label{eq:EOBr_star}
\dfrac{dr_*}{dr}=\left(\dfrac{B}{A}\right)^{1/2} \ ,
\end{equation}
where we recall that $B=D/A$. One then needs to replace the $r$-conjugate
momentum $p_r$ by the $r_*$-conjugate momentum $p_{r_*}$, such that $p_{r_*}dr_*=p_r dr$,
i.e. $p_{r_*} = \left(A/B\right)^{1/2}p_r$. The reason for using this
transformation is that $p_{r_*}$ has a finite limit when $r$ tends to the
zero of $A(r)$ (``$\nu$-deformed effective horizon''), while  $p_r$ diverges there.
Actually, as we stop evolving the dynamics around the $\nu$-deformed light ring
(i.e. before reaching the zero of $A(r)$), this change of variables is not really necessary.
It is, however, convenient because it magnifies the radial axis in a crucial region,
and  prevents any excessive growth of the radial momentum during the plunge.
[Let us mention in passing that the same kind of coordinate has been
used in~\cite{Allen:1997xj} for studying gravitational 
perturbations of non-rotating relativistic stars.]

Neglecting (as it is consistent in a 3PN correction term) the square of the
factor $B/A$ entering the  $z_3 p_r^4/r^2$ term, this leads to the
following form for the effective Hamiltonian
\begin{equation}
\hat{H}_{\rm eff}\equiv \sqrt{ p_{r_*}^2 + A\left(1+\dfrac{p_{\varphi}^2}{r^2}\
+z_3\dfrac{p_{r_*}^4}{r^2} \right)} \ .
\end{equation}
Hamilton's equations for $(r,\varphi,p_{r_*},p_{\varphi})$ then read
\begin{align}
%\label{eob:1}
%\dfrac{dr_*}{dt}         &= \dfrac{p_{r_*}}{ \nu\hat{H}\hat{H}_{\rm eff} } \ , \\
%
\label{eob:1}
\dfrac{d\varphi}{dt}     &= \dfrac{A p_\varphi}{\nu r^2\hat{H}\hat{H}_{\rm eff}} \equiv \Omega\ , \\
\label{eob:2}
\dfrac{dr}{dt}           &= \left(\dfrac{A}{B}\right)^{1/2}\dfrac{1}{\nu\hat{H}\hat{H}_{\rm eff}}\left(p_{r_*}+z_3\dfrac{2A}{r^2}p_{r_*}^3\right) \ , \\
\label{eob:3}
\dfrac{dp_{\varphi}}{dt} &= \hat{\cal F}_{\varphi} \ , \\
\label{eob:4}
\dfrac{dp_{r_*}}{dt}     &= -\left(\dfrac{A}{B}\right)^{1/2}\dfrac{1}{2\nu
  \hat{H}\hat{H}_{\rm eff}} \nonumber \\
&\times\left\{A'+\dfrac{p_\varphi^2}{r^2}\left(A'-\dfrac{2A}{r}\right)+z_3\left(\dfrac{A'}{r^2}-\dfrac{2A}{r^3}\right)p_{r_*}^4
  \right\} \ ,
\end{align}
where $A'=dA/dr$. 
In these equations the extra term $\hat{\cal F}_\varphi$ represents the
non conservative part of the dynamics, namely the radiation reaction force. 

During the quasi-circular {\it inspiral}, a rather accurate expression 
for $\hat{\cal F}_\varphi$ is the following Pad\'e-resummed 
form~\cite{Damour:1997ub}
\begin{equation} \label{FK}
\hat{\cal F}_{\varphi}^K \equiv \dfrac{{\cal F}_{\varphi}^{K}}{\mu} =
-\dfrac{32}{5}\nu
\Omega^{7/3}\dfrac{\hat{f}_{\rm DIS}(v_{\Omega};\nu)}{1-v_{\Omega}/v^{\rm
    DIS}_{\rm pole}} \ ,
\end{equation}
which is expressed in terms of the PN ordering parameter  
$v_{\Omega} \equiv \Omega^{1/3}$. In this expression, the function 
$\hat{f}_{\rm DIS}$ denotes the ``factored flux function'' 
of Ref.~\cite{Damour:1997ub}, scaled to the Newtonian 
(quadrupole) flux [see Eq.~(4.6)-(4.8) there].

Ref.~\cite{Buonanno:2000ef} assumed that the analytical continuation of the
expression~(\ref{FK}) might still be a sufficiently accurate
description of radiation reaction effects during the {\it plunge}.
On the other hand, the authors of Ref.~\cite{Damour:2006tr} pointed out
that the expression~(\ref{FK}) assumed the continued validity of the usual
Kepler law~\footnote{modulo a factor $\psi$ taken into account below.}
 $\Omega^2r^3=1$ during the plunge.[This is why we label the expression
 (\ref{FK}) with a superscript $K$, for Kepler.] They, however, emphasized that
  the Kepler combination $K=\Omega^2 r^3$ significantly deviates from one after
the crossing of the LSO, to become of  order of 0.5 at the (effective)
light ring. Ref.~\cite{Damour:2006tr} went on to argue for a different
expression for the radiation reaction, say 
$\hat{\cal F}_{\varphi}$ (without any superscript),
that {\it does not assume} Kepler's law. This new expression reads 
\begin{equation}\label{F}
\hat{\cal F}_{\varphi} \equiv \dfrac{\cal{F}_{\varphi}}{\mu}= -\dfrac{32}{5}\nu \Omega^5 r^4_{\omega}
\dfrac{\hat{f}_{\rm DIS}(v_\varphi;\nu)}{1-v_\varphi/v_{\rm pole}^{\rm DIS}} \ ,
\end{equation}
where $v_{\varphi}=\Omega r_{\omega}$ and $r_{\omega}=r\psi^{1/3}$ where
the function $\psi$ is defined as in Eq.~(22) of 
Ref.~\cite{Damour:2006tr}\footnote{The quantity $r_{\omega}$ is introduced 
to simplify some expressions, because it is such that the  Kepler-looking 
law $\Omega^2r_\omega^3=1$  holds, without correcting factor, during the 
inspiral (i.e. above the LSO).}.
Note that the essential difference between the two possible expressions for
the radiation reaction is that $\hat{\cal F}_{\varphi}^K \propto \Omega^{7/3}$,
while $\hat{\cal F}_{\varphi} \propto \Omega^{5}r^4$. 
See Ref.~\cite{Damour07a} (notably Fig.~2 there) for a detailed comparison
of these two analytical representations of radiation reaction to ``exact''
numerical results during the plunge.
In both possible expressions for the radiation reaction, our current ``best estimate'' of
$\hat{f}$ is obtained by Pad\'e approximating the currently most complete 
post-Newtonian results, namely the 3.5PN 
ones~\cite{Blanchet:2001aw,Blanchet:2004ek,Blanchet:2004bb}.

 In the forthcoming analysis we shall compare (and contrast) the
relative dynamics and the related  final black hole spin
using various PN-accuracies for the EOB dynamics 
(2PN Hamiltonian + 2.5PN radiation reaction, versus  
3PN Hamiltonian + 3.5PN radiation reaction), as well as 
the two different expressions for radiation reaction 
briefly discussed above.

%--------------------------------------------------------------------------
%
%               FIG1: The maximum of orbital frequency
%
%--------------------------------------------------------------------------
%-------------------------------------------------------------------------
%----------------
\section{Results}
\label{sec:results}
%------------------
The computation of the relative dynamics needs two separate steps: (i) to initialize
the system (\ref{eob:1})-(\ref{eob:4}) and (ii) to integrate it in time.
The initial condition for the relative dynamics is given in a standard way, notably by
specifying a non-zero initial value for $p_{r_*}$ according to the PN order that is 
being used. Our implementation follows Eqs.~(4.16)-(4.21) of 
Ref.~\cite{Buonanno:2000ef} and Eqs.~(4.8) and (4.10) of Ref.~\cite{Buonanno:2005xu}
[see also Eqs.~(9)-(13) of Ref.~\cite{Nagar:2006xv}] 
and does not need to be discussed explicitly here. Let us only mention 
that, for the initial relative separation that we shall take, namely 
$r_0=10$, the leading post-adiabatic approximation is sufficient for 
getting a smooth quasi-circular inspiral (without noticeable eccentricity). 

The orbital frequency $\Omega$ develops a maximum at approximately 
the location of the last unstable EOB circular orbit defined by the condition 
$(A(r)/r^2)'=0$. As already mentioned above, in the test-mass limit ($\nu\ll 1$), this condition 
defines the {\it light ring} $r=3$. When $\nu\ne 0$, we shall refer to the solution of
$(A(r)/r^2)'=0$ as the $\nu$-deformed light ring (LR): $r_{\rm LR}(\nu)$.
 
 In the $\nu\ll 1$ limit, it was realized long ago~\cite{Davis:1971gg} that 
the crossing of the light ring by a test particle corresponds 
to triggering the black hole quasi-normal modes. For related reasons 
(discussed in~\cite{Damour07a}), in the comparable-mass case, the crossing of the
$\nu$-deformed light ring corresponds to an abrupt change of description:
before this crossing one can still describe the two black holes as two point masses
with EOB relative dynamics, while after this crossing one can replace the
binary system by a single distorted black hole (as in the 
 ``close-limit'' approximation of colliding black holes~\cite{Price:1994pm}).
In other words, the EOB approach estimates the full waveform by {\it matching}
at $r_{\rm match} \simeq r_{\rm LR}$ the inspiral + plunge waveform computed
from the EOB dynamics to a superposition of quasi-normal modes (QNMs) describing 
the ringdown of the final distorted black hole. 

%--------------------------------------------------------------------------
%
%               FIG2: angular momentum prediction
%
%--------------------------------------------------------------------------
\begin{figure}[t]
\begin{center}
\includegraphics[width=85 mm, height=75 mm]{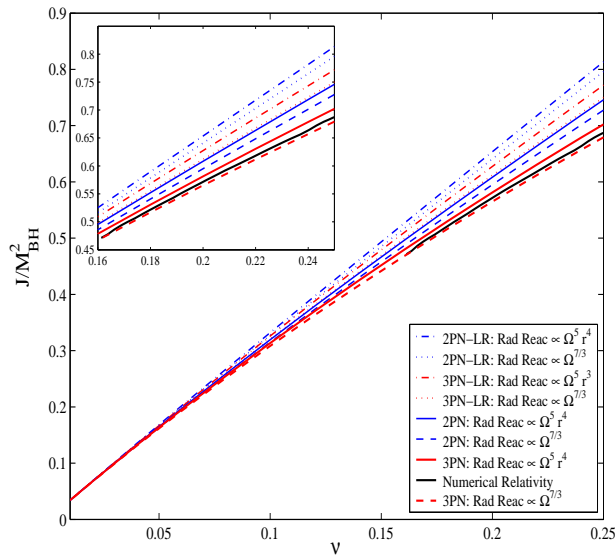}
\caption{\label{fig:fig2} Dimensionless spin parameter of the final black hole:
convergence of Effective-One-Body estimates towards Numerical Relativity ones
as one increases the post-Newtonian accuracy and takes into account the
ringdown losses. See text for discussion.}
\end{center}
\end{figure}
%========================================================================
%
%         Table 1: explicit comparison with Jena data
%
%------------------------------------------------------------------------
\begin{table}[t]
  \caption{ \label{tab:table1}A sample of the numerical data of
    Fig.~\ref{fig:fig2}. From left to right the columns report: the symmetric 
    mass ratio $\nu$, the final dimensionless angular momentum $\hat{a}^{\rm
    NR}$ from Ref.~\cite{Gonzalez:2006md}, 
    and our best estimates (with 3PN+3.5PN dynamics) with 
    $\hat{\cal F}_{\varphi}^K$ ($\hat{a}^K_{\rm BH},M_{\rm BH}^K/M$)  and with 
    $\hat{\cal F}_{\varphi}$ ($\hat{a}_{\rm BH},M_{\rm BH}/M$) . }
  \begin{ruledtabular}
    \begin{tabular}{cccccc}
      $\nu$  & $\hat{a}^{\rm NR}$ & $\hat{a}^K_{\rm BH}$ & $M_{\rm BH}^K/M$ 
      & $\hat{a}_{\rm BH}$ & $M_{\rm BH}/M$ \cr
      \hline
      \hline
      0.25   &   0.6871    & 0.6793 & 0.9555 & 0.7023  & 0.9589\\ 
      0.2402 &   0.6641    & 0.6575 & 0.9582 & 0.6792  & 0.9614\\
      0.2227 &   0.6248    & 0.6181 & 0.9631 & 0.6373  & 0.9659\\ 
      0.2015 &   0.5753    & 0.5687 & 0.9686 & 0.5850  & 0.9710\\
      0.1825 &   0.5281    & 0.5231 & 0.9732 & 0.5370  & 0.9751\\
      0.1613 &   0.4713    & 0.4706 & 0.9778 & 0.4819  & 0.9794\\
      0.14   &      /      & 0.4160 & 0.9821 & 0.4251  & 0.9834\\
      0.12   &      /      & 0.3631 & 0.9858 & 0.3700  & 0.9867\\
      0.10   &      /      & 0.3082 & 0.9891 & 0.3134  & 0.9898\\
      0.08   &      /      & 0.2514 & 0.9920 & 0.2549  & 0.9925\\
      0.06   &      /      & 0.1925 & 0.9946 & 0.1946  & 0.9949\\
      0.04   &      /      & 0.1312 & 0.9968 & 0.1322  & 0.9969\\
      0.02   &      /      & 0.0672 & 0.9986 & 0.0675  & 0.9986\\
      0.01   &      /      & 0.0341 & 0.9994 & 0.0341  & 0.9994
    \end{tabular}
  \end{ruledtabular}
\end{table}
%--------------------------------------------------------------------------------

 Refs.~\cite{Buonanno:2000ef,Buonanno:2005xu} then estimated 
 the mass and angular momentum of the final black hole by the (EOB) energy and angular
momentum of the binary system at the matching point\footnote{Actually,
Ref.~\cite{Buonanno:2005xu} used as end point of the EOB evolution a point
determined by the breakdown of 
 certain adiabatic conditions.} $r_{\rm match}$.
 This procedure, with the choice  $r_{\rm match} = r_{\rm LR}(\nu)$, gives the
 following estimate of the mass and spin parameter of the final black hole
\begin{equation}\label{LR}
M_{\rm BH} \simeq H_{\rm LR}  \qquad\qquad 
\hat{a}_{\rm BH} \simeq \dfrac{P_{\varphi}^{\rm LR}  }{H_{\rm LR}^2} \ .
\end{equation}
This {\it leading-order} estimate of $\hat{a}_{\rm BH}(\nu)$
is plotted for different post-Newtonian approximations 
in Fig.~\ref{fig:fig2} 
and compared with the numerical relativity data of Ref.~\cite{Gonzalez:2006md} 
(black solid line). The labelling of these leading-order ``light ring''
analytical estimates
indicates both the choice of the PN accuracy for the dynamics (``2PN-LR'' or ``3PN-LR''),
and the choice of a specific radiation reaction expression, namely Eq.~(\ref{FK})
(`Rad Reac $\propto \Omega^{7/3}$') or Eq.~(\ref{F}) (`Rad Reac  $ \propto \Omega^{5}r^4$').

It is evident from Fig.~\ref{fig:fig2} that these ``LR'' approximations overestimate
 the ``actual'' result. It is also clear that using a 3PN-accurate dynamics,
instead of a 2PN one,  goes in the good direction.

As we were mentionning in the introduction, the basic physical ingredient which is
lacking in these simple-minded LR estimates is the loss of angular momentum
during the ringdown phase (the importance of this loss was first emphasized in
\cite{Baker:2002qf}). Here we shall use, as next approximation beyond the
leading-order estimates (\ref{LR}), an approximate treatment of the ringdown losses
of angular momentum (and energy) which takes 
advantage of the corresponding results in the small mass case.
Recently, by combining EOB and   
Regge-Wheeler-Zerilli techniques, Ref.~\cite{Nagar:2006xv} computed the gravitational
wave signal (decomposed in multipoles) from the late inspiral, plunge and
ringdown in the small $\nu$ limit. In this calculation the dynamics of the source
($r(t), \varphi(t)$)
is determined together with the waveform $\Psi_{\ell m}(u)$, and this allows one to relate the
retarded time $u$, in terms of which the waveform is expressed, to the dynamical time $t$,
entering the equations of motion (\ref{eob:1})-(\ref{eob:4}). In particular, this
allows one to define precisely the retarded time, say $u_{\rm LR}$,
 corresponding to the crossing of the
light ring, and thereby the part of the waveform which corresponds to the ringdown (in
the EOB sense). It was also verified in  Ref.~\cite{Nagar:2006xv} that the
ratio  $\Psi_{\ell m}(u)/\mu$ had a universal limit as $\nu \to 0$. By integrating
over $u \geq u_{\rm LR}$ quadratic expressions in $\Psi_{\ell m}(u)/\mu$ and its
derivative~\cite{Nagar:2006xv}, and summing over the
various multipoles up to $\ell=4$, 
 one obtains the
following numbers for the energy and angular momentum losses during the
ringdown, in the limiting case $\nu \ll 1$:
$(M/\mu^2) \bar{E}^{\rm ringdown}  \simeq 0.2448$ and  
$\mu^{-2} {\bar{J} } ^{\rm ringdown}  \simeq 1.3890 $ .

Though we know (and discuss below) that the proportionality of these
losses to $\mu^2 \propto \nu^2$ is only strictly valid when $\nu \ll 1$,
our proposal here is to define a {\it next-to-leading order approximation}
 (beyond the leading-order one (\ref{LR})) by simply {\it rescaling} those losses
 (proportionally to $\nu^2$)\footnote{We use here the fact that
such simple-minded $\nu$ rescalings have been found to be (surprisingly) rather accurate, see e.g.
\cite{smarr1978} for the energy loss in the head-on collision of two black holes,
and its comparison to the test-mass limit \cite{Davis:1971gg}. }
 up to any finite value of $\nu$. In other words, let us consider
 that the angular momentum
and energy losses due to the ringdown phase ($u \geq u_{\rm LR}$) are approximately given by
\begin{align}
\bar{E}_{\rm scaled}^{\rm ringdown}(\nu)  \simeq 0.2448 \, \nu^2 M \ , \\
\bar{J}_{\rm scaled}^{\rm ringdown}(\nu)  \simeq 1.3890 \, \nu^2 M^2 \ ,
\end{align}
so that we can define the following next-to-leading order approximations to
 the mass and spin  of the final Kerr black
hole:
\begin{align}
M_{\rm BH}(\nu) &\equiv H_{\rm LR} - \bar{E}_{\rm scaled}^{\rm ringdown}(\nu) \ , \\
J_{\rm BH}(\nu) &\equiv P_{\varphi}^{\rm LR}-\bar{J}_{\rm scaled}^{\rm ringdown}(\nu), \\
\label{NLO}
\hat{a}_{\rm BH}(\nu)&\equiv\dfrac{J_{\rm BH}(\nu)}{M_{\rm BH}^2(\nu)} \ .
\end{align}

 We plot in Fig.~\ref{fig:fig2} the next-to-leading order estimate (\ref{NLO}) for 
 both  choices of radiation reaction and for different post-Newtonian
approximations (they are labelled by, e.g., `2PN: Rad Reac $\propto \Omega^{7/3}$').
It is clear on Fig.~\ref{fig:fig2} that these  next-to-leading order estimates 
are  closer to the numerical relativity ones. It is also clear that the use of the
3PN-accurate dynamics improves significantly the results, compared
to the 2PN-accurate case. Actually, we see that the 3PN+3.5PN order next-to-leading
estimates  (\ref{NLO}) closely  ``bracket'' the numerical 
relativity result. Using the (originally proposed) ``Kepler-type'' radiation
reaction (\ref{FK}) leads to a spin parameter which is slightly smaller than the numerical 
relativity one, while using the more recently proposed radiation
reaction (\ref{F}) leads to a spin parameter which is slightly larger than the
numerical one. Note also that our analytical predictions are computed
for all values of $\nu$ within the full range $0 < \nu < 0.25$. This allows us
to {\it predict} (e.g. by using the 3PN: Rad Reac $\propto \Omega^{7/3}$ curve, or some
combination of the two 3PN curves) the angular 
momentum of the final black hole for the range of  values $\nu < 0.16$, which 
has not been explored (yet) by numerical simulations ( but in which
 the analytical approximation should be rather reliable).
  For the 
sake of comparison, we list in Table~\ref{tab:table1}
our {\it best} (bracketing) 3PN+3.5PN numbers for the dimensionless
spin parameter, together with a selected sample of the numerical
relativity data. We mention, in passing, that had we estimated the 
spin parameter as the ratio 
$(P_{\varphi}^{\rm LR}-\bar{J}_{\rm scaled}^{\rm ringdown})/H_{\rm LR}^2$
(i.e. neglecting the energy loss due to ringdown), and worked at the 3PN level
with our (a priori preferred) radiation reaction $\hat{\cal F}_{\varphi}$,
we would have obtained values even closer to
the numerical relativity ones: e.g.
$\hat{a}(0.25)\simeq 0.6804$, $\hat{a}(0.1825)\simeq
0.5281$,and $\hat{a}(0.1613)\simeq 0.4757$. This (probably
partly accidental) agreement illustrates the fact that our
current (approximate) analytical framework has
already captured most of the correct physics, and that small
variations in the implementation of the EOB approach can probably
lead to an excellent agreement with numerical relativity results.

%----------------
\section{Conclusions}
\label{sec:conclusions}
%------------------

In conclusion, we have improved previous analytical estimates, derived within the
Effective One Body (EOB) framework, of the final spin
of an unequal-mass coalescing  black-hole binary. Our improvements consist in taking a
more complete account of the most relevant physics: we have used higher PN accuracy,
we considered several ways of modelling the radiation reaction during the plunge, and, most
crucially, we took into account the angular momentum and energy lost to gravitational
radiation during the ringdown. Our final results differ by less than $2\% $ from
the recent numerical relativity estimates of \cite{Gonzalez:2006md} 
(see Table ~\ref{tab:table1}). This nice agreement shows, in our opinion, the ability
of the EOB approach to capture, qualitatively and quantitatively,
 the essential physics of the plunge and merger
of black-hole binaries. 

We did not try here to further reduce the small remaining
difference between analytical and numerical
results\footnote{Nor do we wish to conclude from Table I that $\hat{\cal F}_{\varphi}^K$ 
is a more accurate expression for the radiation reaction than $\hat{\cal F}_{\varphi}$.
One needs to consider the effect of higher PN contributions before reaching any
conclusion.}. Our goal here was mainly to exhibit, in a simple case study,
how the inclusion of more and more physical effects in the EOB approach led
to a nice, monotonic convergence towards a numerical relativity result.
In separate investigations \cite{Damour07a,Golm} we shall illustrate how
the EOB framework can also nicely converge towards the gravitational
waveform. The aim of these studies is to understand which parts
of the physics included in the EOB method must be more precisely modelled
to yield accurate  representations of the various physical observables
of merging binary black holes. Indeed, the general philosophy of the
EOB approach is that this (resummed perturbative) analytical framework contains several
{\it flexibility parameters} which can be determined by fitting
EOB predictions to some non-perturbative data, such as numerical
relativity simulations, or, possibly, actual observational data.
An example of a flexibility parameter is the coefficient $a_5(\nu)$
parametrizing presently uncalculable 4PN (or higher) 
additional contributions $+ a_5(\nu)/r^5 + \ldots$ 
to the crucial ``radial potential'' $A(r)$ in the effective metric
Eq.(\ref{eq:eff_metric}). Ref.~\cite{Damour:2002qh} exemplified how
 the parameter $a_5$ (such that $a_5(\nu) = a_5 \nu$) could be
 fitted to numerical data (for initial configurations).
 
We are aware of the rather coarse nature of the approximation used
above for estimating the ringdown losses. In fact, the EOB
approach itself provides a better, and more consistent, way of
estimating these losses. Indeed,  by matching, at $r_{\rm match} \simeq r_{\rm LR}$,
 a post-Newtonian improved 
 plunge waveform ($\ell=2$, $m=\pm 2$) (from the EOB 3PN+3.5PN dynamics) to a 
superposition of quasi-normal modes (QNMs) of a Kerr black hole of mass and angular momentum 
given by the energy and angular momentum of the relative dynamics at the 
$\nu$-deformed EOB light ring, we can {\it analytically} determine the amplitude of the ringdown
waveform. Then, from this waveform we can {\it analytically} estimate, within the EOB
approach, the ringdown losses. In view of the many delicate issues connected
with this matching procedure (see \cite{Buonanno:2006ui,Pan:2007nw,Damour07a}),
we leave to future work a detailed discussion. Let us, however, quote some preliminary
results that we have obtained. In the 
$\nu=1/4$ case, and considering, for simplicity, only the
contribution of the quadrupole ($\ell=2$, $m= \pm 2$) part of the waveform,
this matching procedure gives $\bar{J}_{\rm
  EOB matched, 2,\pm 2}^{\rm ringdown}(1/4)\simeq 0.0899$
for the angular momentum carried away by $\ell=2$, $m=\pm 2$ gravitational
waves (GW) during ringdown.
This happens to be  in good agreement (relative difference $\lesssim 4\%$)
with our naively scaled estimate:$\bar{J}_{\rm scaled}^{\rm ringdown}(1/4)\approx 0.0868$.
However, this good agreement is partly accidental. Several complicated effects go
here in various (probably compensating) directions: (i) the matching can 
under- or over-estimate the amplitude of the ringdown signal, (ii) the exact
scaling with $\nu$ is not exactly proportional to $\nu^2$, (iii) in the  $\nu=1/4$ case,
the GW signal is dominated by the quadrupole $\ell=2$, $m=\pm 2$, (iv) for smaller values of $\nu$
 (when the reflection symmetry is lost) the higher multipoles provide significant
 contributions, and, indeed, our scaling estimate was based on the small $\nu$
limit and included  all the multipoles 
up to $\ell=4$. [As an example of the importance of higher multipoles
for smaller $\nu$'s let us mention that we indeed found, for $\nu =0.1$
a quadrupole-only matched value of   $\bar{J}_{\rm EOB matched, 2, \pm 2}^{\rm ringdown}(0.1)\simeq
0.00631$ which is roughly  half  the scaled 
value $\bar{J}_{\rm scaled}^{\rm ringdown}(0.1)\approx 0.0139$]. 

As a further remark, we point out that  our analytical results above
suggest that a simple quadratic expression $\hat{a}_{\rm fit}(\nu)=a_1\nu+a_2\nu^2$ 
should provide a reasonable fit to the data. And indeed, fitting 
 the data of~\cite{Gonzalez:2006md} with this 
function, provides a good fit\footnote{While writing up our results for publication
we became aware of the recent work \cite{Berti:2007fi} where similar fits are
advocated (for similar reasons to those discussed here). Note, however, that~\cite{Berti:2007fi}
does not focus as we do on the losses during the ringdown phase, but during the
entire plunge + merger phase (post-LSO). As explained above, it is our use of the
analytical EOB description which allows us to define precisely the ``ringdown phase''.}
when  $a_1=3.27690$ and $a_2=-2.11405$. 
Moreover, this fit is found to stay close to our best analytical predictions
for the range $\nu < 0.16$ not covered by numerical simulations.
For instance, For $\nu=0.1$, 
this gives $\hat{a}_{\rm fit}(0.1)\simeq 0.3066$, to be compared with,
say, $\hat{a}^K(0.1)\simeq 0.3082$, while for $\nu=0.01$,
this gives $\hat{a}_{\rm fit}(0.01)\simeq 0.0325$ 
to be compared with,
say, $\hat{a}^K(0.01)\simeq 0.0341$. Note also that the
analytically expected value in the $\nu \to 0$ 
limit is given by  $\hat{a}  = \sqrt{12} \; \nu + O(\nu^2)
=  3.4641 \, \nu + O(\nu^2)$, where the analytical value $\sqrt{12}$
for the $a_1$ coefficient derives from the well known
specific angular momentum of a test particle at the LSO.
The slight difference between $a_1^{\rm analytical}=\sqrt{12}= 3.46410 $
and  $a_1^{\rm fit}=3.27690$ is probably due to the deviations
from analyticity in $\nu$, as $\nu \to 0$, implied by the appearance
of strange fractional powers of $\nu$ (integer powers of $\nu^{1/5}$)
during the transition between the LSO and the plunge, see 
Ref.~\cite{Buonanno:2000ef}.

Let us finally mention that, though we focussed here on nonspinning binaries,
we intend to study, within the EOB approach, the dimensionless spin
parameter of {\it spinning} black-hole binaries. Indeed, not only is
the function  $\hat{a}(m_1,m_2, {\bf S}_1, {\bf S}_2)$  a useful
diagnostics for comparing analytical and numerical results, but it  
has also an important physical meaning. If the cosmic censorship
conjecture is correct, this function should always stay smaller than 1, 
even if the individual spins take their maximum Kerr values 
$S_1 = m_1^2$, $S_2 = m_2^2$. Both leading-order EOB analytical 
results~\cite{Damour:2001tu,Buonanno:2005xu}, and recent full-scale  
numerical~\cite{Campanelli:2006uy} results have indicated that 
this is indeed the case. It would be, however, quite interesting 
to improve the EOB estimates beyond the leading order, and to 
compare them in detail with numerical results.

\acknowledgments
We are grateful to J.~Gonzalez, B.~Br\"ugmann, M.~Hannam, S.~Husa and U.~Sperhake 
for sharing with us their numerical data for Fig.~\ref{fig:fig2}. 
The numerical computations have been carried out by  means of the commercial 
software MATLAB$^{\rm TM}$. AN is grateful to K.~Kokkotas and ILIAS for support.
AN thanks IHES and AEI for hospitality during the inception and the development 
of this work.

\end{document}